\documentclass[useAMS,usenatbib,twocolumn,preprintnumbers,nofootinbib]{revtex4}
\usepackage{epsfig}
\input psfig.sty
\pdfoutput=1

\usepackage{graphicx}
\usepackage{dcolumn}
\usepackage{bm}
\usepackage{epsfig}
\usepackage{color}
\usepackage{graphicx}
\usepackage{dcolumn}
\usepackage{bm}
\usepackage[english]{babel}

\begin{document}

\title{On the cosmic distance duality relation and the strong gravitational 
 lens power law density profile}

\author{F. S. Lima$^{3}$}\email{felixsilva775@live.com}

\author{R. F. L. Holanda$^{1,2,3}$}\email{holanda@uepb.edu.br}

\author{S. H. Pereira$^4$}\email{s.pereira@unesp.br}

\author{W. J. C. da Silva$^1$}\email{williamjouse@fisica.ufrn.br}

\affiliation{$^1$Departamento de F\'{\i}sica, Universidade Federal do Rio Grande do Norte,Natal - Rio Grande do Norte, 59072-970, Brasil}

\affiliation{$^2$Departamento de F\'{\i}sica, Universidade Federal de Campina Grande, 58429-900, Campina Grande - PB, Brasil}

\affiliation{$^3$Departamento de F\'{\i}sica, Universidade Federal de Sergipe, 49100-000, Aracaju - SE, Brazil}

\affiliation{$^4$ Departamento de F\'isica, Faculdade de Engenharia de Guaratinguet\'a, Universidade Estadual Paulista (UNESP), 12516-410, Guaratinguet\'a - SP, Brasil}


\begin{abstract}
Many new strong gravitational lensing (SGL) systems have been discovered in the last two decades with the advent of powerful new space and ground-based telescopes. The effect of the lens mass model (usually the power-law mass model) on cosmological parameters constraints has been performed recently in literature. In this paper, by using SGL systems and Supernovae type Ia observations, we explore if the power-law  mass density profile ($\rho \propto r^{-\gamma}$) is consistent with the cosmic distance duality relation (CDDR), $D_L(1+z)^{-2}/D_A=\eta(z)=1$, by considering different lens mass intervals. {  It has been obtained that the verification of the  CDDR validity is significantly dependent on lens mass interval considered: the sub-sample with $\sigma_{ap} \geq 300$ km/s (where $\sigma_{ap}$ is the lens apparent stellar velocity dispersion) is in full agreement with the CDDR validity, the sub-sample with  intermediate $\sigma_{ap}$ values ($200 \leq \sigma_{ap} < 300)$ km/s is marginally consistent with $\eta=1$  and, finally, the sub-sample with low  $\sigma_{ap}$ values ($\sigma_{ap} < 200$ km/s) ruled out the CDDR validity with high statistical confidence. Therefore, if one takes the CDDR as guarantee, our results suggest that using a single density profile is not suitable to describe lens with  low  $\sigma_{ap}$ values  and it is only an approximate description to lenses with intermediate mass interval. }

\end{abstract}


\maketitle
  
\section{Introduction}

The Etherington reciprocity theorem \cite{Eth} establishes an important relation between the luminosity distance $D_L$ and angular diameter distance $D_A$ of an object, the so-called cosmic distance duality relation (CDDR). While $D_L$ is a distance measurement associated with the decrease of the brightness of an object, $D_A$ is associated with the measurement of its angular size. With the suppositions that photons follow null geodesics and its number is conserved along cosmic evolution, the CDDR is expressed as 
\begin{equation}
{D_L(z) \over  D_{A}(z) (1 + z)^{2}} \equiv \eta(z)=  1\;,\label{CDDR}
\end{equation}
where $z$ is the redshift of the object. 

The above relation plays an essential role in observational cosmology and for this reason several recent works propose new methods to test the validity or departures of the CDDR, which could indicate the possibility of systematic errors in observations or even the necessity of new physics \cite{BASSET} if deviations are persistent. Some tests are model dependent \cite{bernardis,avgoustidis2010,uzan2005,avgousti2012,more2016,hol2011,piazza2016}, which put more restrictive limits on possible deviations from (\ref{CDDR}), while several other are model independent \cite{hol2010,PUXUN,liwu2011,gon2012,hol20121,hol20122,costa20151,hol20161,jailson2011,meng2012,yang2013,ELLIS2013,jhingan,LIOO,holper2017,holbarros2016,shafieloo2013,rana,rana20162,linli2018,hol20172,Hol,hol2017bla,Lia,Fu2019,Ruan2018,Hol2019cps,Chen2020,daSilva2020,Xu2020,Lin2020,Zheng2020,Zhou2021} and use only astrophysical quantities. 

The basic approach to test CDDR has been to consider several deformed expressions or parameterizations for $\eta(z)$ in (\ref{CDDR}). Some examples are: (I) $\eta(z)=\eta_0$, (II) $\eta(z)=\eta_0+\eta_1 z$, (III) $\eta(z)=\eta_0+\eta_1 z/(1+z)$,  (IV) $\eta(z)=\eta_0+\eta_1 \ln(1+z)$ and (V) $\eta(z) = (1+z)^{\eta_1}$, where $\eta_0$ and $\eta_1$ and are constant to be constrained with different observational data, as angular diameter distance (ADD) of galaxy clusters, galaxy cluster gas mass fraction (GMF), type Ia supernovae (SNe Ia), strong gravitational lensing (SGL), gamma ray bursts (GRB), radio compact sources (RCS), cosmic microwave background (CMB) radiation, baryon acoustic oscillations (BAO), gravitational waves (GW), Hubble parameter (H(z)) data, etc. Deviation from the CDDR are given by $\eta_0\neq 1$ and $\eta_1 \neq 0$. As a basic result from several works on literature, the CDDR validity has been verified at least within $2\sigma$ C.L. for the parameters $\eta_0, \,\eta_1$. 

Taking, for instance, some recent works, Zhou et al. \cite{Zhou2021} have used parameterization (II) with $\eta_0=1$ and found that $\eta_1=0.047^{+0.190}_{-0.151}$ at $1\sigma$ C.L. by using the latest data set of 1048 SNe Ia and 205 strong gravitational lensing systems.  In \cite{Xu2020} the parameterizations (II), (III) and (IV) have been tested with BAO and SNeIa  and CDDR remains valid at $2\sigma$ C.L.. In \cite{Lin2020} it is used strong lensed gravitational waves as probes to test CDDR with parameterizations (II) and (III) and they found that the model can be constrained at 1.3\% C.L. for (II) and at 3\% C.L. for (III). Zheng et al. \cite{Zheng2020} have tested the parameterizations (II), (III), (IV) and (V) with multiple measurements of quasars acing as standard probes and found that CDDR is valid at least at $1\sigma$ C.L.. Reference \cite{Hol2019cps} presents in Table 1 a comparison for different tests with the parameterizations (II) and (III), involving combinations of ADD, GMF, SNe Ia and H(z) data. Particularly, using SNe Ia plus 61 X-ray measurements from galaxy cluster and Sunyaev-Zel’dovich effect the authors found $\eta_1=0.05 \pm 0.07$ for the first parameterization and $\eta_1=0.09 \pm 0.16$ for the second, both at $2\sigma$ C.L. with $\eta_0=1$.

An interesting method to compare the above parameterization functions was presented recently by da Silva et al. \cite{daSilva2020}. By using Bayesian inference the authors have tried to answer the question of which $\eta(z)$ should be more viable. They use diameter angular distance from galaxy cluster plus SNe Ia and GMF plus SNe Ia data. Tables II and III of \cite{daSilva2020} show the constraint in the parameters for the parameterizations (I), (II) and (III). The statistical constraints on all the functions implied that the CDDR remains valid at $1\sigma$ C.L. in the analyses by using SNe Ia and galaxy cluster GMF and at 2$\sigma$ C.L. when diameter angular from galaxy clusters and SNe Ia data were considered. The Bayesian inference analysis has shown that the model (III) with $\eta_0=1$ was weakly favored in the CDDR test considering the diameter angular from galaxy clusters and SNe Ia data. On the other hand, considering the galaxy cluster GMF and SNe Ia, all three parameterizations had inconclusive or moderate evidence. In both methodologies, model (I) with $\eta_0=1$ is in agreement within 2$\sigma$ C.L..

The data coming from SGL systems have proved to be very useful to perform tests of the CDDR. The main point is that these kind of sources make it possible to test the CDDR at higher redshifts ($z>2$). SGL is an effect arising from Einstein's theory of general relativity and actually is an important tool to probe the nature of dark matter in the universe. The Einstein radius, $\theta_E$, is an important quantity that characterises the lens and it is known that it varies with cosmological model through the ratio of the angular diameter distances between observer/source and lens/source. It occurs when the observer (o), the source (s) and the lens (l) are well aligned \cite{Sch}.  Under the simplest assumption of a SIS describing the mass distribution in lenses, the Einstein radius is given by:
\begin{equation}
 \label{thetaE_SIS}
{{\theta}_E}=4{\pi}{\frac{D_{A_{ls}}} {D_{A_s}}}{\frac{{\sigma}^2_{SIS}} {c^2}},
\end{equation}
where $c$ is the speed of light, $D_{A_{ls}}$ and $D_{A_{s}}$ are the angular diameter distances between (l)-(s) and (s)-(o), respectively, and $\sigma_{SIS}$ is the dispersion velocity due to lens mass distribution. 

The $\theta_E$ measurements plus SNe Ia data have been used recently to perform constraints on $\eta(z)$ functions from (\ref{CDDR}). In \cite{Hol} the authors consider the parameterizations (II) and (III) above with $\eta_0=1$ and use 95 data points from 118 SGL systems obtained by \cite{Cao},  580 SNe Ia from Union2.1 \cite{Suz}, the latest results from {\it Planck} collaboration \cite{Ade,Planck2018}  and WMAP9 satellite \cite{WMAP9} to put limits on $\eta_1$. The results are displayed in Table I of \cite{Hol} and the validity of the CDDR was verified at $2\sigma$ C.L. in all cases. In \cite{hol2017bla} the four parameterizations (II), (III), (IV) and (V) have been analysed with SNe Ia, GRB distance modulus data and two different approaches to describes the mass distribution in SGL, namely a singular isothermal spherical (SIS) model and a more general power-law (PLAW) index model. For the parameterization (II) with $\eta_0=1$ the authors found $\eta_1=0.15 \pm 0.13$ for the SIS model and $\eta_1=0.00 \pm 0.10$ for the PLAW model. For the parameterization (III) with $\eta_0=1$ they found $\eta_1=-0.18^{+0.45}_{-0.65}$ (SIS) and $\eta_1=-0.36^{+0.37}_{-0.42}$ (PLAW). For the parameterization (IV) with $\eta_0=1$ they found $\eta_1=0.22^{+0.40}_{-0.32}$ (SIS) and $\eta_1=-0.10 \pm 0.24$ (PLAW). For the parameterization (V) they found $\eta_1=0.27^{+0.22}_{-0.38}$ (SIS) and $\eta_1=-0.16^{+0.24}_{-0.51}$ (PLAW). All the results are for $2\sigma$ C.L.. In \cite{Lia} the parameterization (II) with $\eta_0=1$ was tested, the authors assumed a flat universe and use 60 SGL-SNe Ia pairs and found a large interval allowed for $\eta_1$, namely $-0.1 \leq \eta_1 \leq 0.8$ at 2$\sigma$ C.L.. In another context, the Ref.\cite{Chen2018} found that the limits on the cosmological parameter $\Omega_M$ were quite weak and biased, and also heavily dependent on the lens mass model when treated the slope of the mass density profile as a universal parameter for all lens galaxies.

As one may see, the analysis of the above models involving SGL indicates that the results are strongly dependent on the density profile describing the mass distribution of gravitational lenses system. The possible evolution of mass density power-law index with redshift for SGL systems has been tested in recent works \cite{Cao2016mnras,Hol2017pj,Amante2019,Chen2018}, and the results suggest the need of treating low, intermediate and high-mass galaxies separately.  By taking the  Planck best-fitted cosmology, the Ref. \cite{Cao2016mnras}
considered SGL observations and relaxed the assumption that stellar luminosity and total mass distribution follows the same power-law. Their results indicated that a model in which mass traces light is rejected at $>$ 95\% C.L.. The authors also found that  the presence of dark matter in the form of a mass component is distributed differently from the light (see also the Ref.\cite{Schwab}).  {   It is worth noting  that recent  cosmological estimates by using SGL systems with $\sigma_{ap} <  210$ km/sec were found to be in disagreement with SNe Ia  and CMB estimates (see Figures 1, 2, and 3 of Ref.~\cite{Amante2019}).}

{  In this paper, we test the CDDR  by using SGL systems and SNe Ia observations and search for some tension between  results considering   SGL system sub-samples differing each other by their lens mass intervals. For this purpose, we use 3 sub-samples from the original 158 SGL systems compiled in the Ref. \cite{Cao} jointly with SNe Ia observations from Pantheon compilation. The analyses are performed by using 133 SGL systems (those with $z < 2.3$, the higher redshift of SNe Ia). These 133 data points are divided into 3 sub-samples according to the lens  stellar apparent velocity dispersion \cite{Cao}: high mass systems ($\sigma_{ap} \geq 300 $ km/s), intermediate mass systems ($200 \leq \sigma_{ap} < 300 $ km/s) and low mass systems ($\sigma_{ap} < 200 $ km/s).  As a basic result, we verify that the CDDR validity depends strongly on the stellar velocity dispersion (or lens mass) interval considered. Particularly, the sub-sample with low  $\sigma_{ap}$ values ($\sigma_{ap} < 200$ km/s) rule out the CDDR validity with high statistical confidence. If the result from this sub-sample is further confirmed by future and better SGL surveys it would bring to light a possible  evidence for new
Physics. On the other hand, if one takes the CDDR as guarantee, these results suggest that using a single density profile, such as $\rho \propto r^{-\gamma}$, is not suitable to describe lens with  low  $\sigma_{ap}$ values  and it is only an approximate description to lenses intermediate mass interval. It is worth to comment that the assumption that the stellar luminosity and total mass distributions follow the same power law is still  assumed. }

This paper is organized as follows: Section II briefly revises the method. In Section III we present the samples used in the analysis. In Section IV we show and discuss the main results of the paper. We summarize the results of the work in Section V.

\section{Method}

We assume a flat universe and the method to test the CDDR validity is based on \cite{Lia}, which does not depend on assumptions concerning the details of cosmological model. We use SGL systems and SNe Ia observations.

From Eq. (2) we define the observational quantity: 
\begin{equation}
\label{D}
D={\frac{D_{A_{ls}}} {D_{A_s}}}=\frac{{\theta}_E c^2}{4{\pi} \sigma^2_{SIS}}.
\end{equation}
In a flat universe, the comoving distance $r_{ls}$ is given \cite{Bal} by
$r_{ls}=r_s-r_l$, and using $r_s=(1+z_s)D_{A_s}$, $r_l=(1+z_l)D_{A_l}$  and $r_{ls}=(1+z_s)D_{A_{ls}}$, we find
\begin{equation}
\label{d2}
D= 1 - \frac{(1+z_l)D_{A_{l}}}{(1+z_s)D_{A_{s}}}.\label{eq6}
\end{equation}
Using (\ref{CDDR}), Eq. (\ref{eq6}) can be written as
\begin{equation}
\label{d3}
\frac{(1+z_s)\eta(z_s)}{(1+z_l)\eta(z_l)}= (1-D)\frac{D_{L_s}}{D_{L_l}}.
\end{equation}
Thus,  with  SGL measurements and knowing the luminosity distances to the lens and source of each system, it is possible to test the CDDR by studying the evolution of $\eta(z)$, for which we consider the following  particular parameterization:

\begin{itemize}
\item  $\eta(z)=1+\eta_0 z$.
\end{itemize}
As one may see, if $\eta_0=0$ the CDDR validity is obtained.

		\begin{figure}[t]
		\centering
		\includegraphics[scale=0.27]{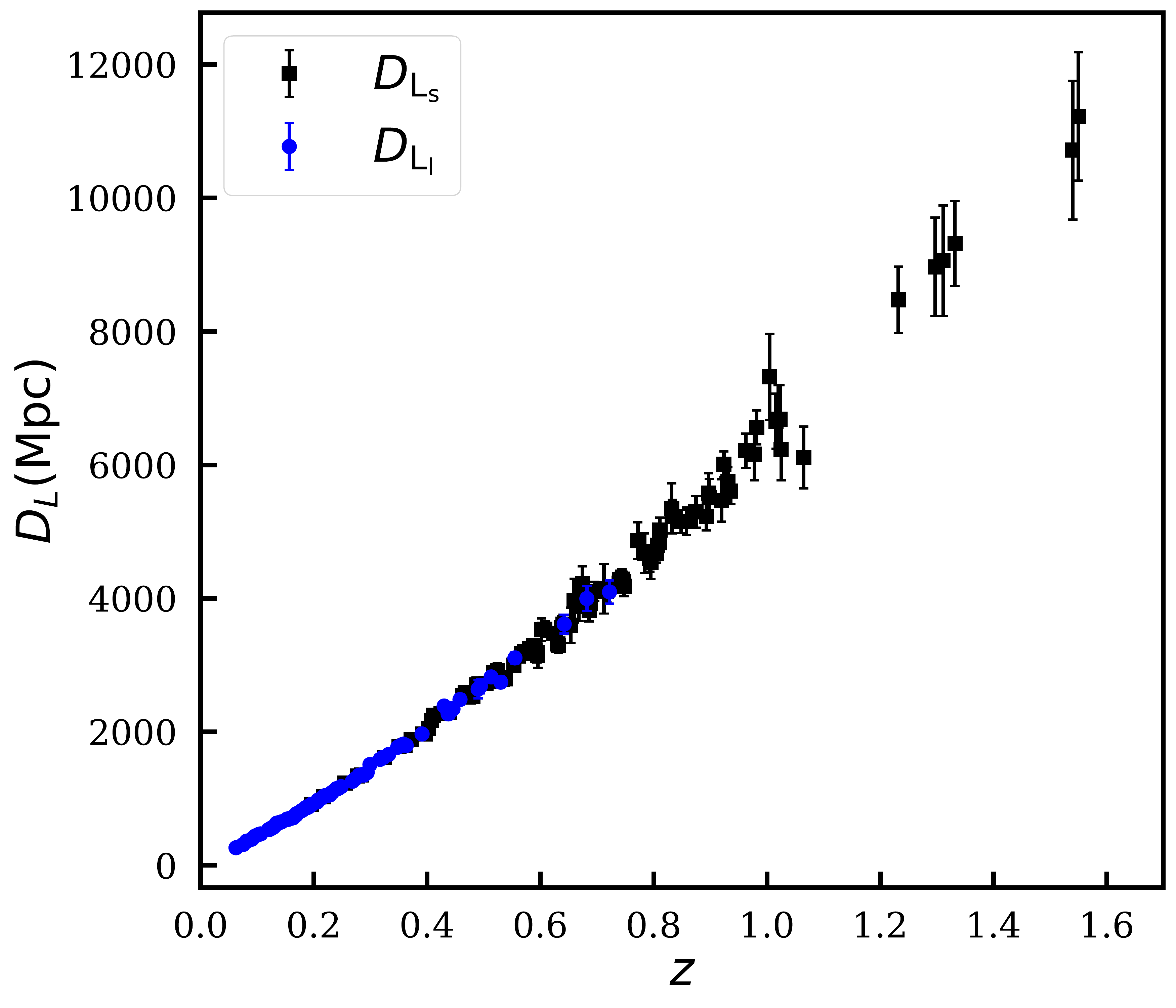}
		\caption{\label{fig:regions2} Luminosity distances  in lens and source of each SGL system obtained from  the Pantheon compilation.} 
		\end{figure}
	
	\begin{figure*}[t]
		\centering
		\includegraphics[scale=0.33]{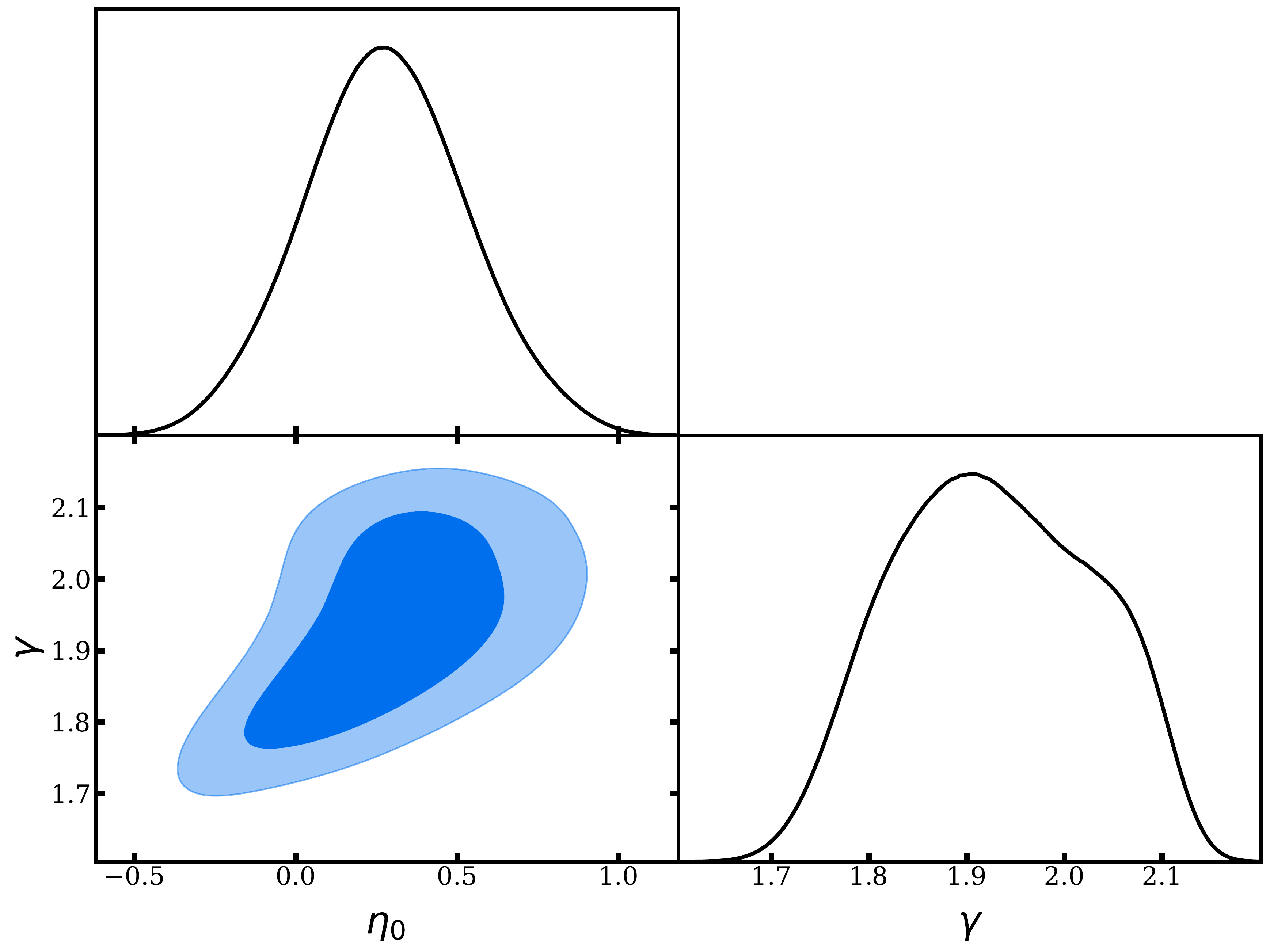}
		\caption{\label{fig:regions2} Results for $\eta_0$ and $\gamma$ parameters by considering the High mass SGL system sub-sample.}
	\end{figure*}

\begin{figure*}[t]
		\centering
		\includegraphics[scale=0.33]{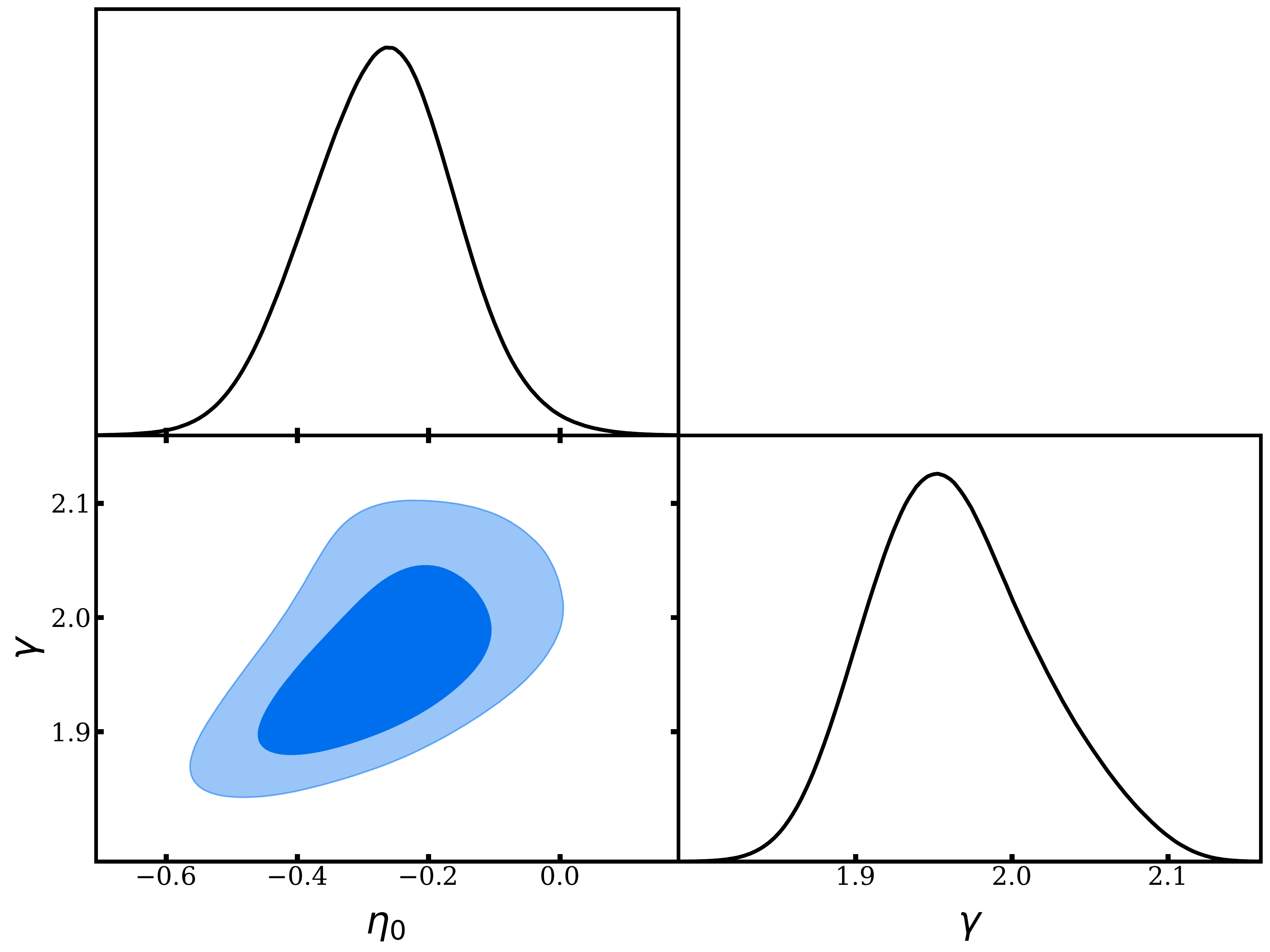}
		\caption{\label{fig:regions2} Results for $\eta_0$ and $\gamma$ parameters by considering the Intermediate mass SGL system sub-sample.}
	\end{figure*}
	\begin{figure*}[t]
		\centering
		\includegraphics[scale=0.33]{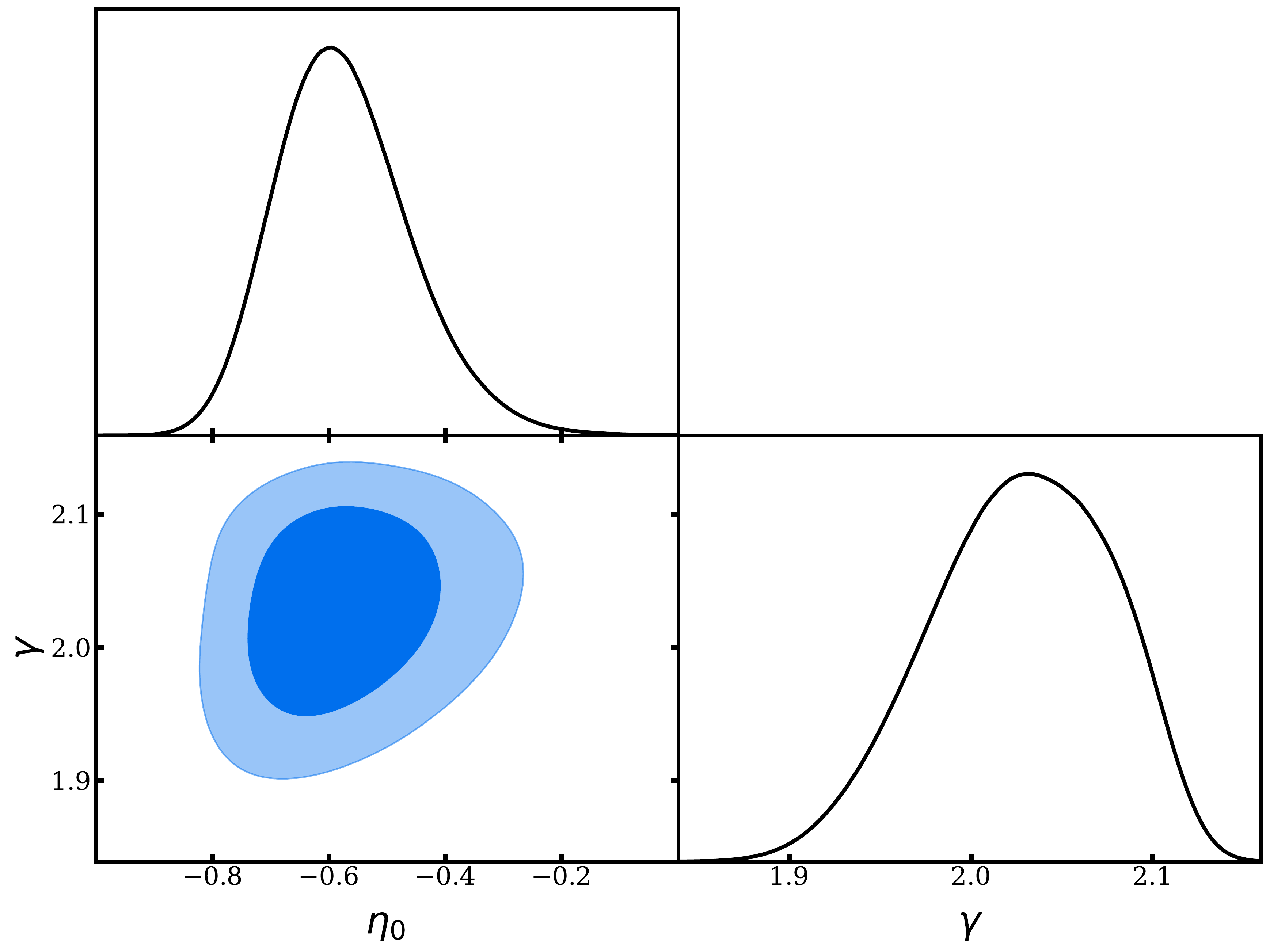}

		\caption{\label{fig:regions2} Results for $\eta_0$ and $\gamma$ parameters by considering the Low mass SGL system sub-sample.}
	\end{figure*}

\begin{table*}[]
\begin{tabular}{lcc}
\hline
$\sigma_{\rm ap} \ [\rm{km/s}]$ & $1 + \eta_0z$               & $\gamma$   \\ \hline
 \hline
$\sigma_{\rm ap} < 200$         &  $-0.575^{+ 0.230}_{ - 0.210}$     & $2.029 \pm 0.092$       \\
$200 \leq \sigma_{\rm ap} < 300$   & $-0.270 \pm 0.220$           & $1.964 \pm 0.104$         \\
$\sigma_{\rm ap} \geq 300$         &  $0.280 \pm 0.510$      &  $1.930 \pm 0.180$ \\
 \hline
\end{tabular}
\caption{\label{table1} $\eta_0$ and $\gamma$ values at $2\sigma$ C.L. for each sub-sample.}
\end{table*}

\section{Samples}

In this work, we use the following data sets:

\begin{itemize}
 
 \item {\it Angular diameter distances}: we consider a specific catalog containing 158 confirmed sources of strong gravitational lensing by the Ref. \cite{Leaf} . This compilation includes 118 SGL systems identical to the compilation of the Ref. \cite{Cao}, which were obtained from SLOAN Lens ACS, BOSS Emission-line Lens Survey (BELLS), and Strong Legacy Survey SL2S, along with 40 new systems recently discovered by SLACS and pre-selected by Ref. \cite{Shu} (see Table I in Ref. \cite{Leaf}). For the mass distribution of lensing systems is considered the so-called power-law model. This one assumes a spherically symmetric mass distribution with a more general power-law index $\gamma $, namely $\rho \propto r^{-\gamma}$ (several studies have shown that slopes of density profiles of individual galaxies show a non-negligible deviation from the SIS \cite{SIS,SIS2,SIS3,SIS4}). In this approach the Einstein radius is:
\begin{equation}
 \label{Einstein} 
\theta_E =   4 \pi
\frac{\sigma_{ap}^2}{c^2} \frac{D_{ls}}{D_s} \left(
\frac{\theta_E}{\theta_{ap}} \right)^{2-\gamma} f(\gamma),\label{thetaE}
\end{equation}
where $\sigma_{ap}$ is the  stellar velocity dispersion inside the aperture of size $\theta_{ap}$ and
\begin{eqnarray} \label{f factor}
f(\gamma) &=& - \frac{1}{\sqrt{\pi}} \frac{(5-2 \gamma)(1-\gamma)}{3-\gamma} \frac{\Gamma(\gamma - 1)}{\Gamma(\gamma - 3/2)}\nonumber\\
          &\times & \left[ \frac{\Gamma(\gamma/2 - 1/2)}{\Gamma(\gamma / 2)} \right]^2.
\end{eqnarray}
Thus, we obtain: 
\begin{equation} 
\label{NewObservable}
 D=D_{A_{ls}}/D_{A_{s}} = \frac{c^2 \theta_E }{4 \pi \sigma_{ap}^2} \left( \frac{\theta_{ap}}{\theta_E} \right)^{2-\gamma} f^{-1}(\gamma).
\end{equation}
For $ \gamma = 2$ we recover the SIS distribution. The relevant information necessary to obtain $D$ in (\ref{thetaE})  can be found in Table 1 of \cite{Cao}.  We marginalize over the $\gamma$ parameter by using the flat prior for $\gamma$: $1.15 < \gamma < 3.5$. The complete data (158 points) is reduced to  133 points whose redshifts are lower than $z = 2.33$ and with the quantity $D$, distance ratio (see next section), compatible with $D = 1$ within 1$\sigma$ C.L. ($D > 1$ represents a non physical region). The 3  sub-samples consist of: 32, 88 and 13 data points with low, intermediate and high $\sigma_{ap}$, respectively. Our compilation contains only those systems with early type galaxies acting as lenses, with spectroscopically measured stellar velocity dispersion, estimated Einstein radius, and both the lens and source redshifts.

\item {\it Luminosity distances}: we use a sub-sample of the latest and largest Pantheon SNe Ia sample in order to obtain $D_L$ of the galaxy clusters. The Pantheon SNe Ia compilation consist of 1049 spectroscopically confirmed SNe Ia covering the redshift range $0.01 <  z < 2.3$ \cite{Pantheon}.  The  sample of $D_L$ is constructed from the apparent magnitude ($m_b$) of the Pantheon catalog by considering $M_b=-19.23 \pm 0.04$ (the absolute magnitude) by the relation
\begin{equation}
    D_L=10^{(m_b - M_b - 25)/5} \mathrm{Mpc}.
\end{equation}
and  $\sigma^2_{D_L}=(\frac{\partial D_L}{\partial \mu} )^2 \sigma^2_{\mu}$. We need luminosity distances to the lens and source of each SGL system. These quantities are obtained as follows: for each one of the 158 SGL systems,
 we carefully select SNe Ia  with redshifts obeying the criteria (I) $|z_{l} - z_{SNe\,Ia}| \leq 0.005$
 and (II) $|z_{s} - z_{SNe\,Ia}| \leq 0.005$.  Finally, we calculate the following weighted average for the distance luminosity selected in each case
\begin{equation}
\begin{array}{l}
\bar{D_L}=\frac{\sum\left(D_{L_{i}}/\sigma^2_{D_{L_{i}}}\right)}{\sum1/\sigma^2_{D_{L_i}}} ,\hspace{0.5cm}
\sigma^2_{\bar{D_L}}=\frac{1}{\sum1/\sigma^2_{D_{L_{i}}}}.
\end{array}\label{eq:dlsigdl}
\end{equation}
After all, we end with a sample containing 133 SGL systems and 133 pairs of luminosity distances (see Fig. 1).

\end{itemize}

\section{Analysis and Results } 
The constraints on the $\eta_0$ parameter are derived by evaluating the likelihood distribution function, ${\cal{L}} \propto e^{-\chi^{2}/2}$, with
\begin{eqnarray}
\chi^{2} & = & \sum_{i}^{N}\frac{\left[\frac{(1+z_{s_i})\eta(z_{s_i})}{(1+z_{l_i})\eta(z_{l_i})}- (1-D_i)\frac{\bar{D_{L_{s_i}}}}{\bar{D_{L_{l_i}}}}\right]^2}{\sigma_i^2},     \end{eqnarray}
{  where $\sigma_i^2$ stands for the statistical errors associated to the $\bar{D_L(z)}$ quantity of the SNe Ia and the gravitational lensing  observations and it is obtained by using standard propagation errors techniques on these quantities.} For the gravitational lensing error one may show that:
\begin{equation} \label{uncertainty}
\sigma_D = D \sqrt{4 (\delta \sigma_{ap})^2 + (1-\gamma)^2 (\delta \theta_E)^2}\;.
\end{equation}
To carry out the statistical analysis, we implement the public package \textsf{MultiNest} \cite{feroz2007, Feroz2019, buchner2014} through the \textsf{PyMultiNest} interface \cite{pymultinest}.

{  In the Figures (2), (3) and (4) we plot the results from our analyses considering the SGL system sub-samples with  high, intermediate and low $\sigma_{ap}$, respectively. The $\gamma$ parameter obtained for each sub-sample is in full agreement each other, being compatible with the SIS model within 1$\sigma$ c.l.. However,  the verification of the  CDDR validity is significantly dependent on lens mass interval considered: the sub-sample with high $\sigma_{ap}$ is in full agreement with the CDDR validity ($\eta_0=0$), the sub-sample with  intermediate $\sigma_{ap}$ values  is marginally consistent with $\eta_0=0$  and, finally, the sub-sample with low  $\sigma_{ap}$ values  ruled out the CDDR validity with high statistical confidence.  Fig.(3) displays this case and, as one may see, the sub-sample with $\sigma_{ap} < 200 $ km/s (32 systems) presents $\eta_0$ values more negative than those from the high and intermediate $\sigma_{ap}$ sub-samples. As commented earlier,  recent  cosmological estimates by using SGL systems with $\sigma_{ap} <  210$ km/sec also were found to be in disagreement with SNe Ia  and CMB estimates (see Figures 1, 2, and 3 of Ref.~\cite{Amante2019}). Then, if one considers the CDDR  valid as guarantee, a simple power law describing the mass density profile of lens is not realistic for low  mass interval and  it is only an approximate description to lenses with intermediate mass interval. In Table I, the $\eta_0$ and $\gamma$ values are shown (with 2 $\sigma$ c.l.). }

\section{Conclusions}

As it largely known, strong gravitational lensing is an important effect arising from Einstein's theory of general relativity, which has played a very important role for testing cosmological models. However, recent papers have suggested the need of treating low, intermediate and high-mass galaxies separately in cosmological analysis (see for instance the Refs. \cite{Chen2018,Amante2019} and references therein). In this paper, we take a look closer at this context by testing the cosmic distance duality relation, $D_L(1+z)^{-2}/D_A=\eta(z)$, with sub-samples of SGL systems that differ from each other by their  stellar velocity dispersion values (or lens mass intervals). We found that  the sub-sample containing SGL systems with $\sigma_{ap} <  200$ km/sec is  inconsistent with the validity of the CDDR (see Table I). On the other hand,  the sub-sample with high $\sigma_{ap}$ ($\sigma_{ap} >  300$ km/sec) was found is in full agreement with the CDDR validity ($\eta=1$) while the sub-sample with  intermediate $\sigma_{ap}$ values  is  marginally consistent with $\eta=1$. This independent result, by using the CDDR context, is in full agreement with other recent results \cite{Chen2018,Amante2019} that also found some tension between cosmological analyses using  SGL systems with $\sigma_{ap} <  200$ km/sec and those from SNe Ia and CMB data. It is remarkable that a local property like mass density profile of SGL systems might be constrained by a global argument provided by the cosmic distance duality relation.

It is worth to comment that as an interesting extension of the present work, one may check the consequences of relaxing the rigid assumption that the stellar luminosity and total mass distributions follow the same power law. Besides, the well-known Mass-sheet degeneracy (see \cite{birrer} and references therein) in the gravitational lens system and its effect on the results on this paper also could be explored.

\section*{Acknowledgments}
RFLH is supported by INCT-A and CNPq (No. 478524/2013-7;303734/2014-0). SHP acknowledges financial support from  {Conselho Nacional de Desenvolvimento Cient\'ifico e Tecnol\'ogico} (CNPq)  (No. 303583/2018-5).

\end{document}